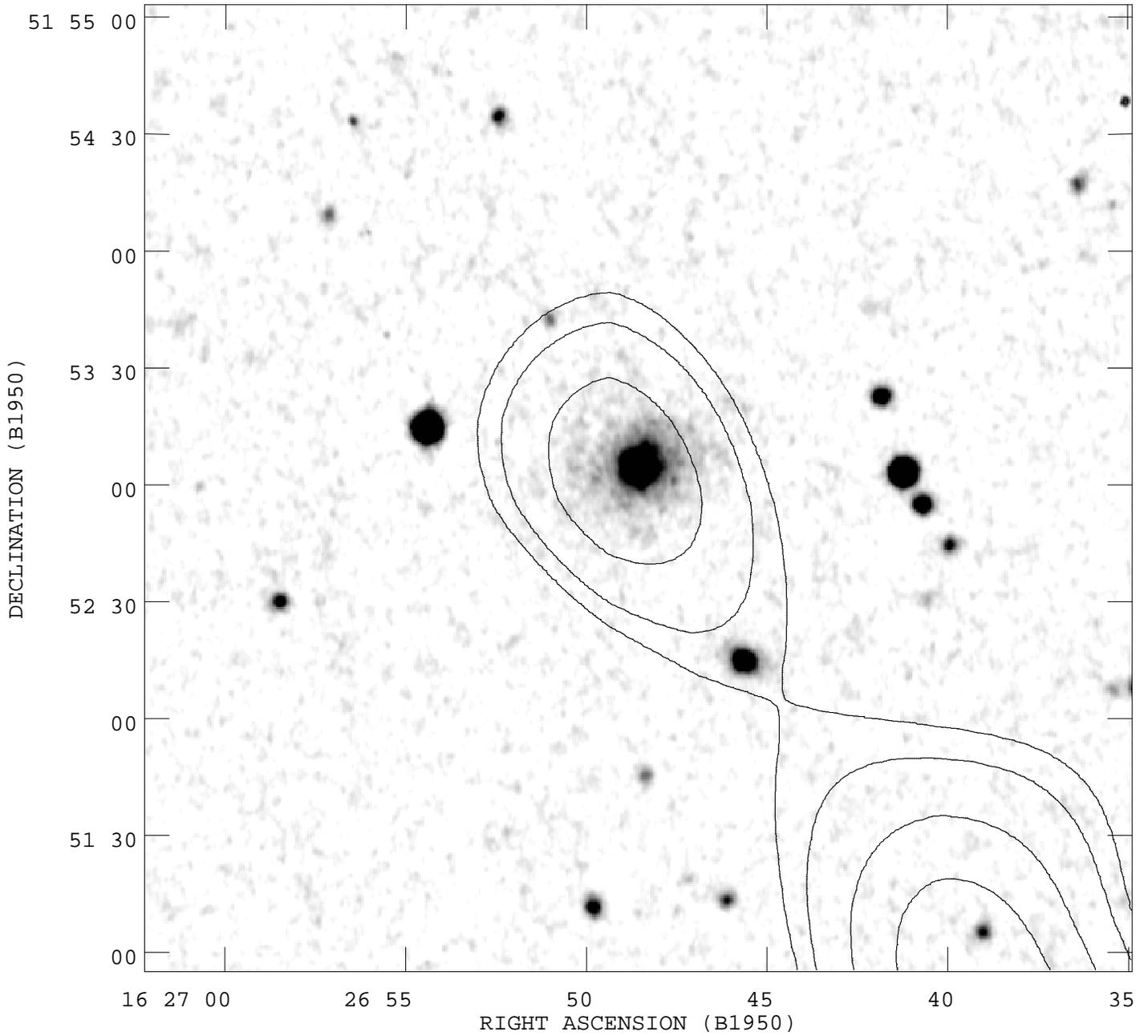

```
Plot file version    3  created 09-DEC-1994 14:31:33
GREY: NONE         GIANT O.TRANS.1
CONT: WN50.254   328.192 MHZ   WN50.254.HGEOM.1
```

Grey scale flux range=    5.5    12.6 Kilo
Peak contour flux =  1.5456E-01 JY/BEAM
Levs =  3.3294E-03 * (  -3.00, 3.000, 4.000,
 6.000, 8.000, 10.00, 15.00, 30.00, 50.00,
 75.00, 100.0, 200.0)

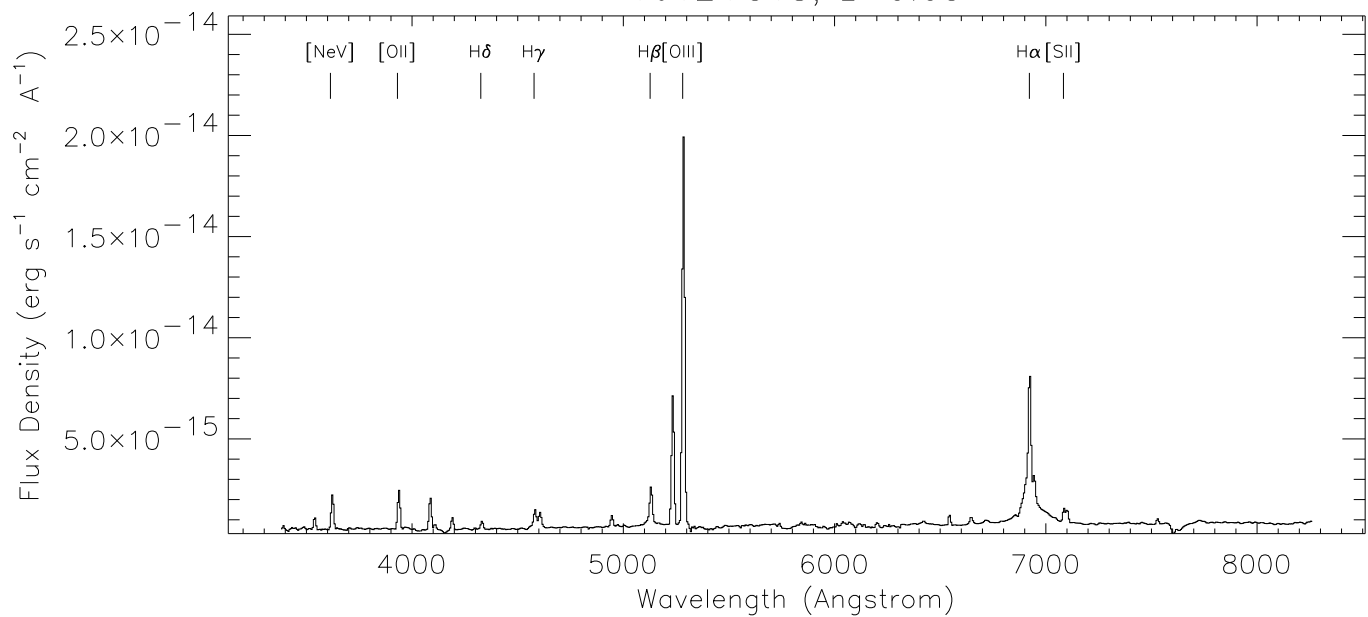

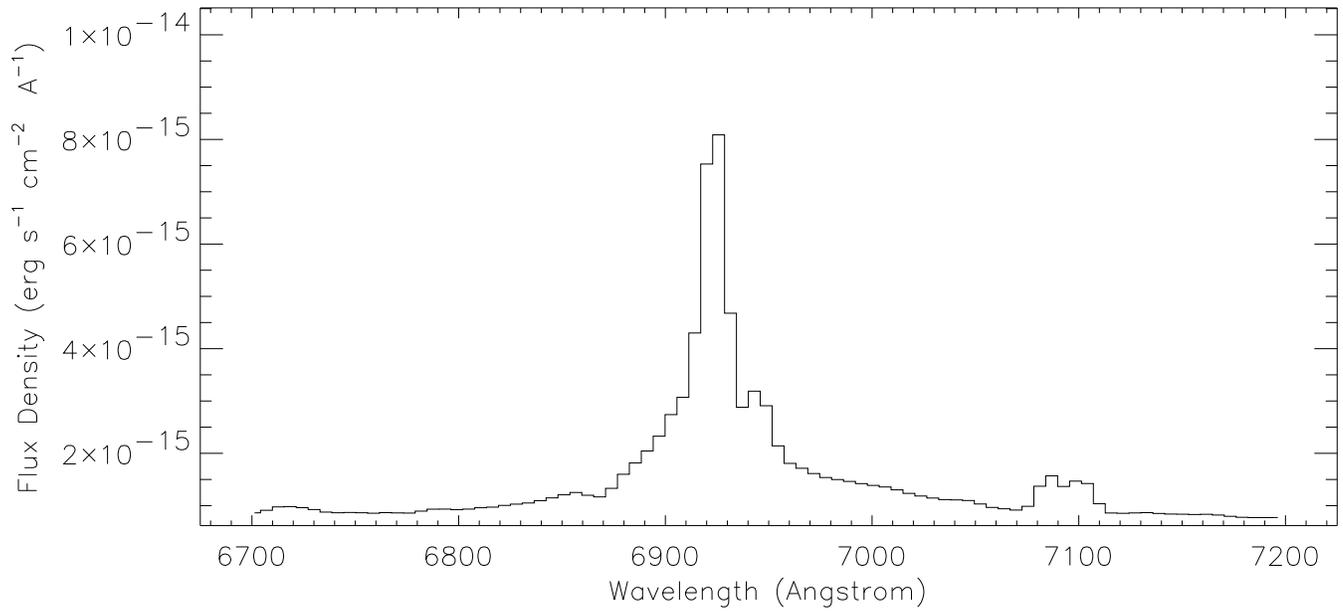

# WN1626+5153: A Giant Radio Galaxy from the WENSS survey

H.J.A. Röttgering,[1,2,3] Y. Tang,[4] M. A. R. Bremer,[1,5] A. G. de Bruyn,[4,6]
G. K. Miley,[1] R. B. Rengelink,[1] M. N. Bremer[1,3]

[1] *Leiden Observatory, PO Box 9513, 2300 RA, Leiden, The Netherlands*
[2] *Mullard Radio Astronomy Observatory, Cavendish Laboratory, Madingley Road, Cambridge, CB3 0HE*
[3] *Institute of Astronomy, Madingley Road, Cambridge, CB3 0HA*
[4] *Radio Sterrewacht, Dwingeloo, Postbus 2, 7990 AA Dwingeloo, The Netherlands*
[5] *Space Research Organisation of the Netherlands, Sorbonnelaan 2, 3584 CA Utrecht, The Netherlands*
[6] *Kapteyn Astronomical Institute, Postbus 800, 9700 AV Groningen, The Netherlands*

6 August 1996

**ABSTRACT**
An FRII radio source with an angular extension of 18.4 arcminutes has been discovered in the Westerbork Northern Sky Survey that is being carried out at 325 MHz. This source is identified with Mrk 1498 at $z = 0.056$ indicating that its projected extension is 1.6 $h_{50}$ Mpc. The optical spectrum of Mrk 1498 has the typical emission lines of a narrow line radio galaxy; H$\alpha$ is the only permitted line that has a faint broad component. The discovery shows that the good sensitivity of WENSS at low frequencies to low surface brightness features at scales of a few minutes, is uniquely suited to provide well defined and relatively large samples of giant radio galaxies. The future definition of such samples will be important for our understanding of giant radio galaxies.

## 1 INTRODUCTION

Radio galaxies with linear sizes exceeding 1.5 $h_{50}^{-1}$ Mpc$^\star$ form the class of giant radio galaxies (GRGs, Saripalli et al. 1986). Their radio power typically range to $10^{25} < P_{408} < 10^{27}$ $h_{50}^{-2}$ W Hz$^{-1}$, just above the transition that distinguishes powerful FRII sources from the low luminosity FRI sources (Fanaroff & Riley 1974).

The aims of studying this class of object are twofold (e.g. Lacy et al. 1993; Subrahmanyan and Saripalli 1993). First, given that their radio properties are so extreme, they are an ideal laboratory for studying the physics of radio sources. Are they so extreme in size because the ratio of their jet power to the density of their environment is so high, or because they are relatively old radio sources in comparison to the bulk of the FRII population? Secondly, these are the only radio sources that probe the intergalactic medium (IGM) over such a large scale. Studying these sources will therefore constrain the structure of the IGM.

We are currently carrying out the Westerbork Northern Sky Survey (WENSS). It will cover the sky north of 30 degrees declination to a ($5\sigma$) flux limit of about 20 mJy at 325 MHz, and will contain some 250,000 radio sources (e.g.

---

$^\star$ We will use a Hubble constant of $H_0 = 50$ km s$^{-1}$ Mpc$^{-1}$, a density parameter of $\Omega = 1$ and $h_{50} \equiv H_0/(50$ km s$^{-1}$ Mpc$^{-1})$.

Bruyn et al. 1994; Röttgering 1995). This low-frequency-radio survey will enable many types of radio source to be selected at about an order order of magnitude deeper in flux density than hitherto was possible.

Studies of giants have been significantly hampered since the known samples are not well defined and contain only a dozen or so objects. Because of its good sensitivity at a low frequency to low surface brightness features at scales of a few arcminutes, the WENSS survey will provide a well defined and relatively large sample of giant radio galaxies. In this letter we report the discovery of a giant from the first WENSS data.

## 2 PROPERTIES OF THE GIANT

The radio source discussed in this paper has been discovered in the WENSS survey maps at 325 MHz. The $uv$ data for the map that contains the large radio source has been taken during the winter of 1992. A contour map of the radio source is presented in Fig. 1. The resolution of the map is $54 \times 68$ arcsec and the rms noise is 3.3 mJy beam$^{-1}$. The radio source has a classical edge brightened double-lobed structure, with an unresolved compact source in the middle, classifying this source as FRII radio source (Fanaroff & Riley 1974). The distance between the location of the max-ima of emission within the two lobes is 18.4 arcminutes. The total flux density of the source is $S_{325} = 1.9$ Jy. The compact source at the centre is unresolved and has a peak flux density of 25 mJy beam$^{-1}$.

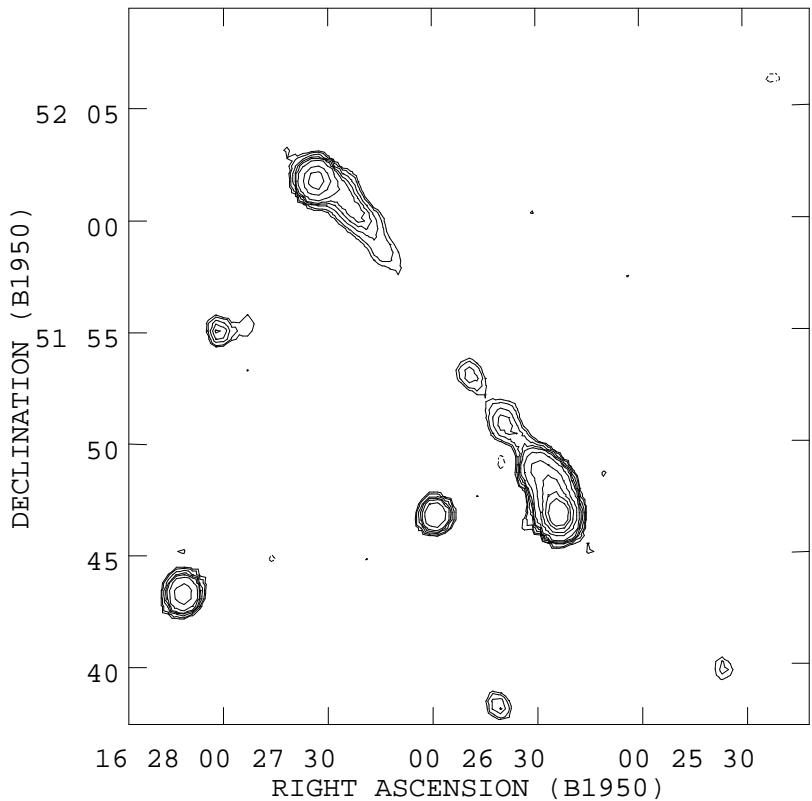

**Figure 1.** A contour map of the giant radio source WN1626+5153 from the WENSS survey. The contour levels are $-3, 3, 4, 6, 8, 10, 15, 30, 50, 75, 100, 200 \times 3.3$ mJy beam$^{-1}$.

We have estimated the total flux density at 4850 MHz using the NRAO Green Bank survey (Condon et al. 1989). A gaussian function was fitted to the northern and southern lobe and to the compact source between the two lobes. Summing the flux of these three components, we find that the total flux at 4850 MHz for the source is $S_{4850} = 320$ mJy, yielding a spectral index $\alpha_{4850}^{325} = -0.66$. The peak flux density at 4850 MHz of the compact component is $S_{4850} = 27$ mJy beam$^{-1}$, giving a spectral index of $\alpha_{4850}^{325} = 0.0$. We therefore assume that this component is the flat spectrum core. Its coordinates as measured from the WENSS map are $16^h 26^m 49.0^s \pm 0.4^s$; $51°53'2'' \pm 4''$ (B1950).

The object was also seen in the 6C survey (151 MHz) as two separate sources (Hales et al. 1990). The lack of spatial resolution and sufficient sensitivity prevented its recognition as a single object. The total flux density of the source at 151 MHz is 3.13 Jy, consistent with the radio spectrum being a simple powerlaw from 151 MHz down to 4850 MHz with a spectral index $\alpha = -0.66$.

The optical position of the Markarian Galaxy 1498 is $16^h 26^m 48.50^s \pm 0.15^s$; $51°53'5'' \pm 1.5''$ (Kojoian et al. 1984). This is well within the errors of the positions of the core of the radio source, strongly suggesting that this radio source is associated with this Markarian galaxy. In Fig. 2 we show in grey scale a part of the Palomar O survey plate around Mrk 1498 as digitised by the APM plate measuring machine (Irwin et al. 1994). The contours on this figure are the central part of the WENSS image of the radio galaxy (e.g. Fig 1). The central objects is Mrk 1498. The extent of Mrk 1498 on the POSS is at least 40 arcsec, with no spiral structure visible indicating that Mrk 1498 is an elliptical galaxy (see also Low et al. 1988). It is associated with IRAS1612+518 at redshift of 0.0547 (e.g. de Grijp et al. 1992). If the identification is correct, the projected size of the radio source is 1.6 $h_{50}^{-1}$ Mpc. Its bolometric infrared luminosity $L(IR) = 10^{10.81} h_{50}^2 L_\odot$ (Low et al. 1988) is typical for a nearby radio galaxy (Knapp et al. 1990).

We have taken an optical spectrum of Mrk 1498 on June, 28, 1995 using the ISIS spectrograph on the 4.2-m William Herschel Telescope (WHT). The object was observed through a slit that had a width of 2 arcsec. The observation was made with the 570 nm dichroic in place; on the blue as well as the red arm we used a Tek chip. The conditions were photometric. The analysis was carried out in a standard way using the iraf twodspec package. In our analysis of the spectrum we have followed Röttgering et al. (1996). The resulting spectrum has a pixel size of 5.7 Å, a wavelength range from 3374 - 8266 Å, a resolution of 12 Å and is shown in Fig. 3. The spectrum resembles that of a narrow emission line radio galaxy (e.g. Osterbrock 1989); the identification of stronger emission lines are indicated in Fig. 3. From the stronger forbidden lines we find that the redshift of Mrk 1498 is $z = 0.056 \pm 0.001$, roughly consistent with the measurements of de Grijp et al. 1992. The (deconvolved)

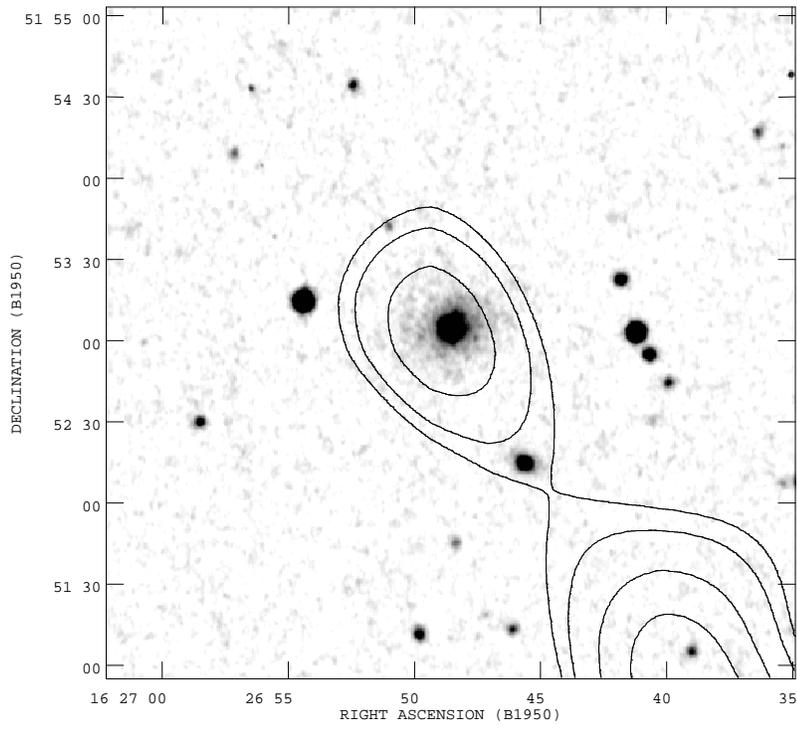

**Figure 2.** A grey scale map of part of the Palomar O survey as digitised by the APM plate measuring machine. The map is centred on the nuclear component of WN1626+5153. The contour levels are from the the WENSS survey (see Fig. 1) and are $-3, 3, 4, 6, 8, 10, 15, 30, 50, 75, 100, 200 \times 3.3$ mJy beam$^{-1}$.

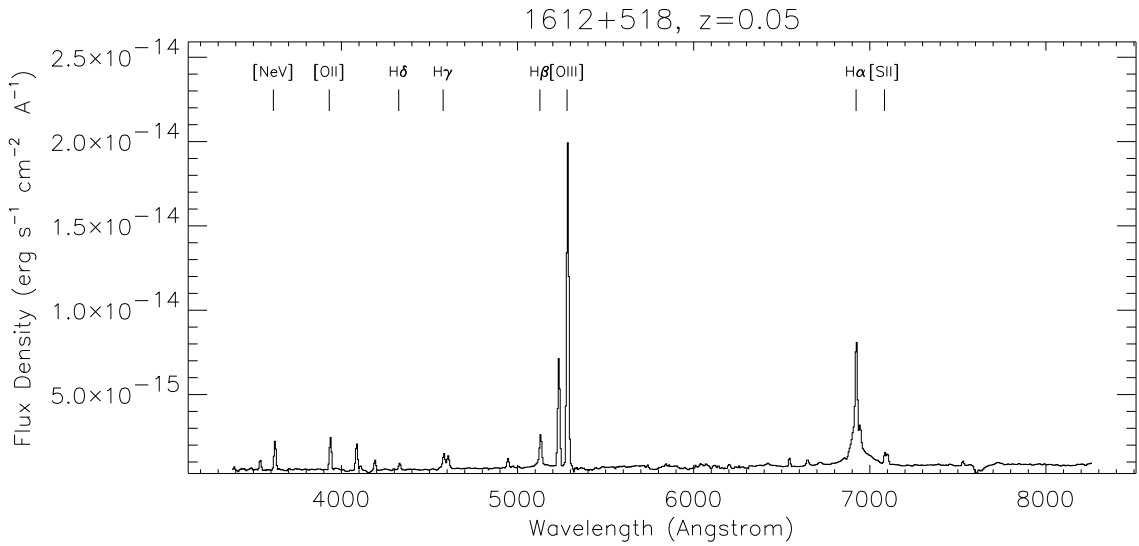

**Figure 3.** The optical spectrum of WN1626+5153, taken with the ISIS spectrograph on the WHT.

4  *H.J.A. Röttgering et al.*

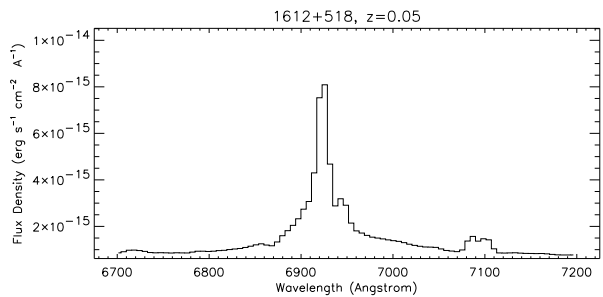

**Figure 4.** Part of the he optical spectrum of WN1626+5153, taken with the ISIS spectrograph on the WHT, showing the region around the Hα emission line.

full width at half maximum of all the lines is typically 400 km s$^{-1}$, typical for a narrow line radio galaxy. The part of the spectrum that contains Hα is shown in Fig 4; Hα is the only permitted line that has a faint broad component (full width at zero intensity of 6500 km $^{-1}$).

On the basis of data with less signal to noise and/or less resolution, the spectrum of this object has been classified as a Seyfert 1 (Markarian et al. 1983; Low et al. 1988; de Grijp et al. 1992). Given our deep spectrum, the classification of Mrk 1498 as a Seyfert 2 is more accurate. Since Mrk 1498 is associated with a powerful radio source, we classify this object as a narrow line radio galaxy.

## 3 DISCUSSION

On the basis of the positional coincidence between the radio nucleus and Mrk 1498, it is very likely that the identification is correct. There are two further arguments that the identification is indeed correct. First, if the host galaxy is not Mrk 1498 but a back-ground galaxy located more than three times further away than Mrk 1498, the size of the radio source would be much larger than the largest radio source known (3C 236 that is 5.7 $h_{50}^{-1}$ Mpc in size, Willis et al. 1974). Secondly, the V-magnitude for the galaxy is 14.9 (de Grijp et al. 1992). Using the relation between redshift and $V$-magnitude that Laing (1980) found for 3CR radio galaxies ($\log z = 0.181 V - 3.965$) we find an estimated reshift of $z = 0.054$, consistent with the measured redshift.

Various models have been proposed to explain the observed differences between radio quasars and radio galaxies. It could be that radio quasars are the progenitors of radio galaxies or that the environment determines whether they becomes a quasar or a radio galaxy (e.g. Norman and Miley 1984). However, most of the observational evidence seem to indicate that the observed differences are mainly due to their orientation relative to the line of sight ("the orientation unification model" e.g. Scheuer 1987; Barthel 1989; Antonucci 1993).

An interesting test for these models is a comparison of the distributions of projected linear radio source sizes for samples of quasars and radio galaxies. Barthel (1989) found that steep spectrum quasars from the 3C catalogue with $0.5 < z < 1.0$ have sizes that are systematically smaller than 3C radio galaxies at similar redshift by a factor of 2.2. This difference has been taken as support for the orienta-tion model, indicating a cone angle of $\sim 45$ deg along which the broad lines of a quasar nucleus can be seem. It is not clear whether this trend also holds at low redshift. It seems that there is a relative shortage of observed steep spectrum quasars at low redshift (Kapahi 1990). This could indicate that the unification scheme breaks down and that there are indeed less quasars than radio galaxies in our local universe. Alternatively, the opening angle of the cone in nearby objects could be less than it is at higher redshifts and therefore less quasars are observed in the nearby universe.

There are two giant radio sources known to have strong broad permitted lines in their optical spectrum. 0309+411 (z = 0.136) is 1.8 $h_{50}^{-1}$ Mpc in size and is identified with a red 18th magnitude galaxy (de Bruyn 1989). 4C74.26 (z=0.136) is 1.6 $h_{50}^{-1}$ Mpc in size and is identified with a stellar object (Riley et al. 1989). More than a dozen or so giant radio sources are known (e.g. Saripalli et al. 1986; Subramanyan et al. 1996). It seems that the existence of now 3 giant radio sources with broad lines is broadly consistent with the orientation unification model.

We should stress however that well defined samples of giants are needed to address this issue properly. Since the WENSS survey has excellent sensitivity at low frequency to faint surface brightness limits and covers a large fraction of the northern sky, it will provide large and well defined samples of giants.

The monochromatic radio luminosity of WN 1626+5153 at 178 MHz is $P_{178} = 3.2 \times 10^{24}$ Watt Hz$^{-1}$ sr$^{-1}$, almost a magnitude below the characteristic radio power $P^*$ that distinguishes FRI and FRII radio sources (Fanaroff & Riley 1974). Studies of samples of giant radio galaxies have been carried out by Saripalli et al. (1986) and Subramanyan et al. (1996). From these 2 samples 2 out of in total 19 FRII sources also have a radio power a magnitude below $P^*$, the remaining 17 all have powers similar or larger than $P^*$. To first order this indicates that $P^*$ is not a strong function of radio source size (cf de Ruiter et al 1990). However, this might well be a selection effect in that sources with radio powers well below $P^*$, were difficult to detect in radio surveys. The significant samples of giant radio galaxies that WENSS will provide, will allow to investigate whether a large fraction of giant radio sources have radio powers well below $P^*$ and that hence $P^*$ is a function of radio source size. Recently, Owen and Ledlow (1994) showed that the FRI/FRII break is a function of both radio and optical luminosity. The behaviour of $P^*$ as a function of these various observational parameters will contrain models of radio source evolution.

## Acknowledgements

We would like to thank Garret Cotter, Richard Saunders and Arno Schoenmakers for useful discussions. and Ignas Snellen for reducing the ISIS spectrum. We acknowledge support from an EU twinning project, funding from the high z programme subsidy granted by the Netherlands Organization for Scientific Research (NWO) and a NATO research grant. The WSRT is operated by the NFRA with financial support from NWO.

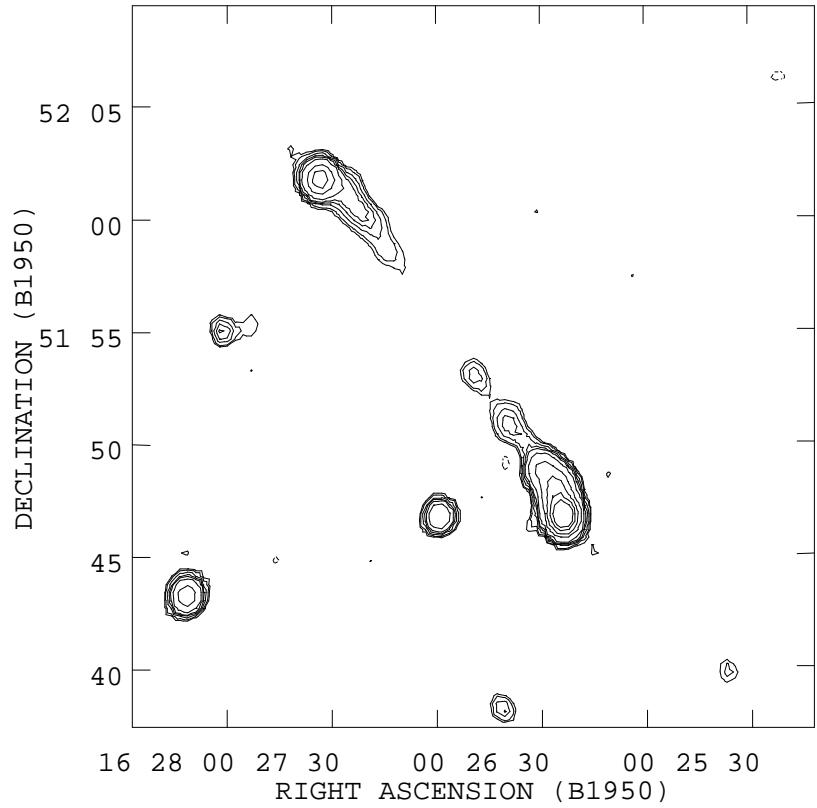